\documentclass{amsart}
\RequirePackage{fix-cm}

\usepackage{wasysym}
\usepackage{amssymb,amsmath,stmaryrd,amsfonts}
\usepackage{mathtools,leftindex}
\usepackage{multirow}

\usepackage{array}

\usepackage{tikz,mwe,pifont}
\tikzset{boximg/.style={remember picture,red,thick,draw,inner sep=0pt,outer sep=0pt}}

\usepackage{natbib}
\usepackage[margin=1.2in]{geometry}

\newcommand{\ie}{\textit{i.e.,}}
\newcommand{\eg}{\textit{e.g.,}}

\newcommand{\norm}[1]{\left\lVert#1\right\rVert}
\newcommand{\dldda}{\norm{\Delta \rm{LD}^{\rm{DA}}}}
\newcommand{\dldfd}{\norm{\Delta \rm{LD}^{\rm{FD}}}}
\newcommand{\dld}{\norm{\Delta \rm{LD}}}
\newcommand{\ld}{\textrm{LD}}

\usepackage{enumitem}   

\usepackage{rotating}

\usepackage[colorlinks,citecolor=blue]{hyperref}

\begin{document}

\title[Improved detection of chaos with Lagrangian descriptors]{
	Improved detection of chaos with Lagrangian descriptors using differential algebra
}

\author[A. C\u{a}liman]{Alexandru C\u{a}liman}
\address{
	Department of Mathematics and naXys, Namur Institute for Complex Systems,
	University of Namur, 61 Rue de Bruxelles, 5000 Namur, Belgium}
\email{alexandru.caliman@unamur.be}

\author[J.\,Daquin]{
	J\'er\^ome Daquin         
}
\address{
	Department of Mathematics and naXys, Namur Institute for Complex Systems,
	University of Namur, 61 Rue de Bruxelles, 5000 Namur, Belgium}
\email{jerome.daquin@unamur.be}

\author[A.-S. Libert]{Anne-Sophie Libert}
\address{
	Department of Mathematics and naXys, Namur Institute for Complex Systems,
	University of Namur, 61 Rue de Bruxelles, 5000 Namur, Belgium}
\email{anne-sophie.libert@unamur.be}

\date{\today}

\begin{abstract}
Lagrangian descriptors (LDs) based on the arc length of orbits previously demonstrated their utility in delineating structures governing the dynamics. Recently, a chaos indicator based on the second derivatives of the LDs, referred to as $\dld$, has been introduced to distinguish regular and chaotic trajectories. Thus far, the derivatives are numerically approximated using finite differences on fine meshes of initial conditions. In this paper, we instead use the differential algebra (DA) framework as a form of automatic differentiation to introduce and compute $\dld$ up to machine precision. 
We discuss and exemplify benefits of this framework, such as the determination of reliable thresholds to distinguish ordered from chaotic trajectories. Our extensive parametric study quantitatively assesses the accuracy and sensitivity of both the finite differences and differential arithmetic approaches by focusing on paradigmatic discrete models of Hamiltonian chaos, namely the Chirikov's standard map and coupled $4$-dimensional variants. 
Our results demonstrate that finite difference techniques for $\dld$ might lead to significant misclassification rate,  up to $20\%$ when the phase space supports thin resonant webs, due to the difficulty to determine appropriate thresholds.
On the contrary, $\dld$ computed through DA arithmetic leads to clear bimodal distributions which in turn lead to robust thresholds. As a consequence, the DA framework reveals as sensitive as established first order tangent map based indicators, independently of the underlying dynamical regime. Finally, the benefits of the DA framework are also highlighted for non-uniform depleted meshes of initial conditions.
\end{abstract}


\maketitle

\tableofcontents

\section{Introduction}\label{sec:intro}
Chaotic dynamics, or sensitivity to initial conditions,  is ubiquitous in nature and observed at different scales, from the Solar system \cite{jLa89} and fluid dynamics \cite{jOt89} to plasma physics \cite{dEs16,dEs18}. 
The long-term unpredictability inherent to chaotic dynamics manifests itself already as the result of ``simple'' mathematical demographic models, such as iterations of quadratic Logistic maps \cite{rMa76}, or small perturbations of integrable one degree-of-freedom Hamiltonians, such as the two-waves Hamiltonian \cite{dEs81}. 
Computational methods  to detect the signature of chaos have catalysed a significant body of literature over the last several decades.\\
This paper is concerned with the recent fast surrogate for chaos detection that follows from Lagrangian descriptors (LDs).  
LDs have been initially introduced for smooth vector fields and popularised for fluid flows \cite{jMa09,cMe10,aMa13}.  They found applications in various fields such as chemistry \cite{gCr15,vKr20}, semiclassical theory \cite{jMo24} or cardiovascular studies \cite{wAb24}, to name but a few. Extension of LDs for discrete mapping have been introduced in \cite{cLo15}. In the following we shall consider regular mapping, 
\begin{align}\label{eq:map}
x'=M(x), \, x \in D \subset {\mathbb{R}}^{d},
\end{align}
and denote by $\{x_{0},x_{1},x_{2},\dots,x_{N}\}$
the orbit obtained by iterating $N$ times the mapping.
The LD at time $N$ associated to Eq.\,(\ref{eq:map}) is given by 
\begin{equation}\label{eq:LD}
    \textrm{LD}(x_0;N)=\sum\limits_{n=0}^{N-1} \| x_{n+1}-x_n \|_{2},
\end{equation}
where $\norm{\bullet}_{2}$ is the Euclidean norm on $\mathbb{R}^{d}$. 
We note that LD satisfies the extended mapping 
\begin{align} \label{eq:GeneralEqDiscreteCase}
 \left\{
 \begin{aligned}
    & x'=M(x),\\
    & \textrm{LD}'=\textrm{LD} + \norm{x'-x}_{2},
\end{aligned} 
\right.
\end{align}
provided $\textrm{LD}(x_{0};0)=0$.  
From the real valued function  $\textrm{LD}: D \subset \mathbb{R}^{d} \to \mathbb{R}_{+}$, 
the scalar
\begin{equation}\label{eq:DLDdef}
    \| \Delta \textrm{LD}(x_0;N) \|=\sum\limits_{i=1}^{d} \big| \partial^2_{i} \mathrm{LD}(x_0;N) \big|,
\end{equation}
turns to be useful to detect and portray chaotic dynamics \cite{jDa22}. We drop from now the final time $N$ in Eq.\,(\ref{eq:DLDdef}).
The indicator is able to portray and recover with great details the topology of phase spaces.
It has been qualitatively benchmarked  on multiple low and higher dimensional discrete and continuous paradigmatic models, such as the Logistic Map, symplectic mappings, the shearless standard map, Hénon-Heiles Hamiltonians and several Escande-Doveil nearly-integrable Hamiltonian systems supporting resonances overlap \cite{jDa22,jDa22-cps2,mHi22,sZi23,sBa24}. As a matter of fact, $\dld$ allows to identify different kinds of dynamics given that the crossing of hyperbolic domains corresponds to sharp increase of its value, in correspondence with the loss of regularity of the LD function.

Visualisation of high-dimensional dynamical systems is a challenge, and the strategy customarily employed consists in computing the considered indicator on specific domains of initial conditions and to encode its magnitude through a heatmap, revealing regions with distinct properties. 
To estimate the second derivatives in Eq.\,(\ref{eq:DLDdef}), the current approach relies on discrete-like Laplacians involving finite differences on fine meshes of initial conditions. In the case of a two-dimensional domain meshed by a regular Cartesian grid $\mathcal{G}$, $\dld$ is approximated at a node $(x,y)$ of $\mathcal{G}$ with the 5-stencil 
$\{(x,y),(x+h,y),(x-h,y),(x,y+h),(x,y-h)\}$  as
\begin{equation}\label{eq:DLDFD}
\begin{split}
\norm{\Delta \textrm{LD}(x,y)}\simeq
h^{-2}\Big(
\big\vert
\ld(x+h,y)+\ld(x-h,y)-2\ld(x,y)
\big\vert 
\\
+ 
\big\vert 
\ld(x,y+h)+\ld(x,y-h)-2\ld(x,y)
\big\vert
\Big),
\end{split}
\end{equation}
where $h$ is the discretisation step size. On the boundary of the domain, Eq.\,(\ref{eq:DLDFD}) is adapted to use either forward or backward differences. Note that we considered here the same grid size $h$ in the $x$ and $y$ coordinates directions, but Eq.\,(\ref{eq:DLDFD}) can be adapted to two different discretisation sizes.

The error of the approximation given by Eq.\,(\ref{eq:DLDFD}) is $\mathcal{O}(h^{2})$. The present work focuses on the implication of such an approximation  when  seeking to distinguish regular from chaotic motions. In fact, in the limit $h \to 0$, it is reasonable to expect that such an approximation of $\dld$ will reflect local properties associated to the initial condition $x_0$. However, typical resolutions of meshes consider $h \sim 10^{-3}$. Given this scale, the fact that 
$\dld$ still reflects local properties associated to $x_0$ is questionable. Our manuscript investigates  this question and presents numerical evidence that this approximation might lead to significant misclassification of ordered and chaotic motions. Our investigations are based on  discrete parametric models stemming from Hamiltonian chaos, namely the standard map and coupled $4$-dimensional versions supporting Arnold's resonant web. 
To reconciliate the situation, we hereby propose 
the use of differential algebra (DA), detailed in \autoref{subsec:DA}, to 
compute $\dld$. The use of DA not only improves the error of Eq.\,(\ref{eq:DLDFD}), but reduces it  to machine precision (using the same initial grid $\mathcal{G}$ where finite differences were initially evaluated). Equipped with $\dldfd$ and $\dldda$, that is $\dld$ computed through finite differences and differential algebra, respectively, we evaluate and compare their  sensitivity in classifying orbits.
The classifications are compared to several well established indicators, that we consider as ground truth. 
Among the large landscape of surrogates that have been developed over the last several decades, we consider in this work deviation vector based methods, including the finite-time Lyapunov Exponent, the fast Lyapunov indicator, the mean exponential growth of nearby orbits, and the smaller alignment index (see the definitions and relevant literature  in~\ref{sec:Appendix2}).

The rest of the paper is structured as follows:
\begin{itemize}
    \item In \autoref{sec:methods}, we review the  central computational methods used here: \autoref{subsec:dld} presents how to select thresholds allowing binary (regular or chaotic) classification with $\dld$; \autoref{subsec:DA} presents the DA framework that will be central for computing  $\dld$ up to machine precision. 
    \item In \autoref{sec:resultsSM}, we  discuss several benefits  in computing the $\dld$ index within a DA framework, such as thresholds determination for chaos classification based on the shape of the  distribution. For this, we present a comprehensive  quantitative performance assessment of the indices against several well established tangent map methods. 
    We investigate the long-term stability of the initial conditions that are inconsistently classified by the finite differences method and the DA arithmetic approaches. We consider two paradigmatic models of Hamiltonian chaos: the 2-dimensional standard map in \autoref{subsec:SM2d} and a 4-dimensional model of coupled standard maps in \autoref{subsec:4d}.
    \item In \autoref{sec:NU}, we assess the performances of the indicators computed on depleted meshes of initial conditions, mimicking the situation of non-uniform fields due to incomplete data sets or corrupted measurements. Such meshes challenge even further the finite differencing approach generally adopted. 
\end{itemize}    
Finally, \autoref{sec:conclusions} closes the paper and summarises our results and conclusions.

%==========
\section{Computational methods} \label{sec:methods}
%==========
This section presents the main computational tools used in this paper. We start by presenting the methodology to set a threshold to binarily classify an orbit as regular or chaotic from its $\dld$ value and then touch on the differential algebra framework.

%==========
\subsection{Threshold selection for orbits' classification with $\dld$
}\label{subsec:dld}
%==========
As previously mentioned, LD assigns to initial conditions the Euclidean lengths of their associated trajectories over a finite-time window.
From the LD index, the diagnostic based on the unmixed second partial derivatives with respect to the initial conditions, denoted $\dld$, has been suggested as global chaos indicator \cite{jDa22}. $\dld$ captures the regularity of the LD metric and is defined, we recall  Eq.~(\ref{eq:DLDdef}), as
\begin{equation}\label{eq:DLD}
    \| \Delta \textrm{LD}(x_0) \|=\sum\limits_{i=1}^{d} \big| \partial^2_{i} \mathrm{LD}(x_0) \big|.
\end{equation}
In order to derive a criterion allowing to classify  in a binary way regular and chaotic orbits (and, beyond, assess the performance of the indicator against existing surrogates, see \autoref{sec:resultsSM}),  it is desirable to find a threshold value $\alpha$ such that at time $N$
\begin{align}\label{eq:ThresholdsDelta}
 \left\{
 \begin{aligned}
    & \norm{\Delta \textrm{LD}(x_{0};N)} \le
    \alpha \Rightarrow \textrm{``The orbit stemming from $x_{0}$ is regular'',}\\
    & \norm{\Delta \textrm{LD}(x_{0};N)} >
    \alpha \Rightarrow \textrm{``The orbit stemming from $x_{0}$ is chaotic''.}
\end{aligned} 
\right.
\end{align}
Such a threshold $\alpha$ is problem dependant and is best determined by following 
a well established procedure that leverages on the shape of the distribution of the considered chaos indicator computed for many initial conditions (this calibration is also used for methods based on tangent maps or variational equations, see \eg \,\cite{jSz05,eSa20}). 
In fact, for a domain dominated by regular orbits, it is observed that the distribution of $\dld$ is clearly unimodal. 
In this case, $\alpha$ is set by taking a value in the right-part of the tail (larger than the mean of the distribution) associated with low probabilities. 
When the domain is dominated by chaotic motions, the situation is similar and $\alpha$ is taken in the left-side (smaller than the mean of the distribution) of the tail  of the unimodal distribution. 
When regular and chaotic orbit coexist in a balanced way, the distribution becomes bimodal and, when the two peaks are clearly identifiable, it has been suggested to take
$\alpha$ as the lowest value between the spikes \cite{mHi22,sZi23,jLo24}. 
Quasi-integrable Hamiltonian dynamics are generally neither uniformly hyperbolic nor integrable, but chaotic and regular structures coexist in an intertwined way. This picture, as it will be illustrated in the subsequent, tends to distort bimodal distributions and introduces skew and fat tails. 
In this case, the determination of $\alpha$ based on the distribution is on shaky ground. The differential algebra framework presented in the next section is much more robust to this threshold selection, as it will be demonstrated in \autoref{sec:resultsSM}.

%===========
\subsection{Differential Algebra for the computation of $\dld$}\label{subsec:DA}

Instead of using finite differences to approximate Eq.\,(\ref{eq:DLD}), we could instead derive the mappings ruling the time evolution of the derivatives involved in $\dld$ and iterate the dynamics accordingly. This process implies the need to explicitly follow, up to the second order, contribution of a basis of tangent vectors. Deriving those equations might be a laborious task, especially for high-dimensional problems.
Here we instead opt for treating the dynamics via the differential algebra (DA) arithmetic. DA allows access to the various derivatives over time automatically, without having to derive and iterate explicitly the mappings they satisfy (indeed, this task is accomplished implicitly by the DA framework, see \cite{jGi23} for flows). 
The generality of the DA methodology also allows to scale up the approach to obtain, if desired, the derivatives of higher orders whatever the dimension of the dynamical system.\\

By a slight abuse of language,  DA refers to the representation and manipulation of sufficiently regular functions through their associated high-order Taylor expansions in a computer environment. It originated from plasma physics \cite{mBe86,mBe87,mBe88} and found in the recent years applications in astrodynamics for accurate nonlinear uncertainty propagation \cite{mVa13}, detection of hyperbolic structures of dynamical systems \cite{dPe15,jTy22}, high-order expansions of flows of ODEs \cite{jGi23} to name but a few.
The mathematical details briefly presented below follow closely \cite{aWi16}, to which we refer for omitted details. Central to DA is the set $\leftindex_{k}{D}_{d}$ of multivariate polynomials over $\mathbb{R}$ of $d$ independent variables and maximal order $k$. This set is endowed  the usual addition, subtraction, scalar multiplication, and multiplication  of polynomials. The  multiplication in turns allows to define division of vectors of $\leftindex_{k}{D}_{d}$. The algebraic structure is further endowed with a differential operator $\partial_{i}$ satisfying Leibnitz rule with respect to the multiplication, defined by
\begin{align}
    \partial_{i}P=\frac{\partial}{\partial x_{i}}P(x), \, x=(x_{1},\cdots,x_{d}) \in \mathbb{R}^d.
\end{align}
The richness of DA resides in interpreting computations over the infinite dimensional space of sufficiently regular functions as operations (implemented into a computer environment) over the corresponding finite dimensional set $\leftindex_{k}{D}_{d}$, by associating to a function $f \in \mathcal{C}^{k}(0)$
its truncated Taylor expansion around $0$ at order $k$, \ie \, a vector of $\leftindex_{k}{D}_{d}$.
Our use of DA arithmetic is motivated by the second derivatives with respect to the initial condition entering Eq.\,(\ref{eq:DLD}). In fact, 
one classical DA application is related to automatic differentiation. In order to obtain the derivatives of a sufficiently regular function $f$ at a point $x_{0}$, it suffices to evaluate 
in DA arithmetic the expression $g(x)=f(x+x_{0})$, corresponding to the Taylor expansion of $g$ around $0$. The Taylor coefficients of this expansion are  the derivatives of $f$ at $x_{0}$ and, by applying successively the differential operators, we are able to retrieve the derivatives from this expansion to floating point error (see omitted details in \cite{aWi16}).\\

\begin{table}
\centering
\begin{tabular}{ll}
\hline
Input: & 1.  The extended mapping of Eq.\eqref{eq:GeneralEqDiscreteCase}. \\
       & 2. The initial condition $(x_0,\textrm{LD}_0)=(x_0,0)$ of the extended map.\\
       & 3. The maximum number of iterations $N$ and the maximum \\ 
       & cut-off value $T$  allowed for $\dldda$.\\ \hline \hline
Step 1 & \textbf{Set} the current iteration $i$ to 
$i \leftarrow 0$ and the stopping flag (SF) \\
    & to $\textrm{SF} \leftarrow 0$.  \\
Step 2 & \textbf{Initialize} \texttt{DACE}: set the number of variables $\textrm{NV}$ to $\textrm{NV}=d$ and    \\
    & the order of expansion (OE) to $\textrm{OE}=2$.\\
Step 3 & \textbf{Initialize} the floating point initial condition $x_0$ in DA arithmetic:\\ 
    & $x_0 \leftarrow x_0+\delta$, $\delta=(\delta_1,\dots,\delta_d)$.\\
Step 4 & \textbf{While} ($i \leq N$ \textbf{AND} $\textrm{SF}=0$) \textbf{Do} \\
       &  \quad \textbf{Evolve} the extended mapping of Eq.\,(\ref{eq:GeneralEqDiscreteCase}) in DA arithmetic.
    \\ 
     & \quad \textbf{Extract} the unmixed second derivatives and form $\dldda$. \\
     & \quad \textbf{If}($\log_{10}(\dldda)>T$) \textbf{Then}  \\
    & \quad \quad \textbf{Set} $\textrm{SF} \leftarrow 1$. \\
    & \quad \textbf{End If}. \\
    & \quad \textbf{Set} $i \leftarrow i+1$.  \\
 &  \textbf{End While} \\
 Step 5 & \textbf{Report} the value of $\dldda$. \\
\end{tabular}
\caption{Procedure  followed for computing $\dldda$. 
The evolution of the initial condition $x_{0}$ and $\textrm{LD}$ are followed in DA arithmetic until the maximum number of iterations $N$ is reached or $\log_{10}(\dldda)>T$. DA computations are performed using the \texttt{DA Computational Engine (DACE)}. The procedure is identical for flows.
} 
\label{table:AlgorithmDA}
\end{table}

To accomplish the task of computing $\dld$ with a DA framework, we used the \texttt{Differential Algebra} \texttt{Core Engine\footnote{Available online: \url{https://github.com/dacelib/dace}.}} (\texttt{DACE}) and  followed the steps detailed in Table \ref{table:AlgorithmDA}.
In the next sections, we are interested in assessing the performances, differences  and sensitivity of both the $\dldfd$ and $\dldda$ as fast chaos indicators. 

%========================
\section{Computational results} \label{sec:resultsSM}
\subsection{Applications to the standard map}\label{subsec:SM2d}
\begin{figure}
    \centering
    \includegraphics[width=0.8\textwidth]{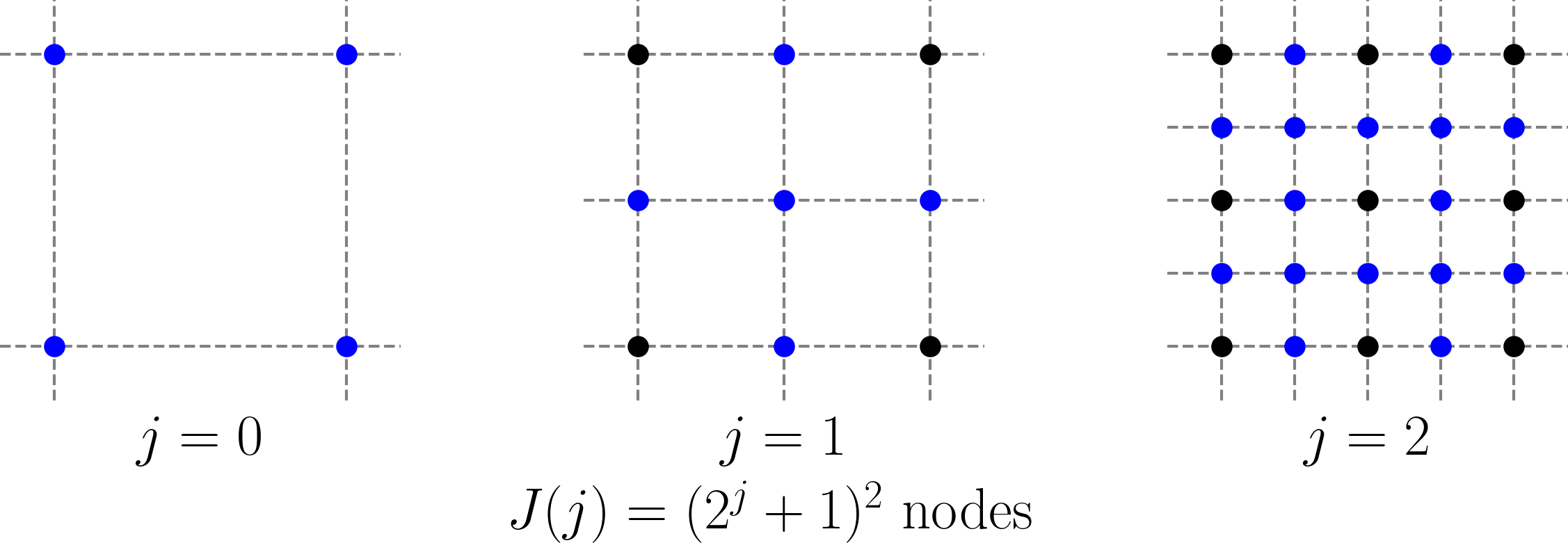}
    \caption{The meshes of the initial domains are obtained by increasingly halving intervals up to a given order $j$. At a given order $j$, nodes' coordinates are given by Eq.\,(\ref{eq:Mesh}) and the mesh contains $J(j)=(2^{j}+1)^{2}$ initial conditions. This process ensures successive meshes, at orders $j$ and $j+1$, to share $J(j)$ nodes, a property that is useful to compare the performances of $\dldfd$ and $\dldda$ to probe the sensitivity of $\dldfd$ with respect to the mesh size. This process is illustrated here for $j=0, 1, 2$. At each iterations, new nodes generated appear in blue. 
    }
    \label{fig:Mesh}
\end{figure}

We start our analysis by probing the phase space of the lifted two dimensional standard map \cite{bCh79,jMe08},
\begin{align}\label{eq:StandardMap2D}
\left\{
    \begin{aligned}
    &x'=x+y', \\
    &y'=y-\frac{k}{2 \pi } \sin(2 \pi x), 
    \end{aligned}
\right.
\end{align} 
that we consider on the unit square and $k \in \mathbb{R}^{+}$ is the nonlinearity parameter.
For $k=0$, $y$ is constant and $x$ evolves linearly with time. For $k > 0$, chaotic motions are known to exist \cite{jMe92,jMe08}.  Instead of a ``linear sampling'', 
we sample a given domain $D=[x_{m},x_{M}]\times[y_{m},y_{M}] \subset [0,1]^{2}$ of initial conditions as a regular Cartesian mesh  depending on a given order $j \in \mathbb{N}$. 
For a given $j$, 
the nodes of the mesh have coordinates $\{e_{1}(p),e_{2}(q)\}$, $(p,q) \in 
 \llbracket 0,2^{j} \rrbracket^{2}$, defined by 
\begin{align}\label{eq:Mesh}
\left\{
    \begin{aligned}
    &e_{1}(p)=x_{m}+\frac{p}{2^{j}}(x_{M}-x_{m}), \\
    &e_{2}(q)=y_{m}+\frac{q}{2^{j}}(y_{M}-y_{m}).
    \end{aligned}
\right.
\end{align}
The grid has $J(j)=(2^{j}+1) \times (2^{j}+1)$ nodes (serving as initial conditions for the mapping) and consecutive maps generated at order $j$ and $j+1$ have $J(j)$ nodes that coincide, as illustrated in the sketch of Fig.\,\ref{fig:Mesh}. This is precisely the property that turns to be useful for our analysis, allowing to compare $\dldfd$ results computed on a mesh at order $j+1$ with $\dldda$ results computed at order $j$.  
Unless otherwise specified, from now on we set $j=9$ meaning that the meshed domain contains $J(9) = 513 \times 513$ initial conditions.\\

\begin{figure}
    \centering
    \includegraphics[width=\textwidth]{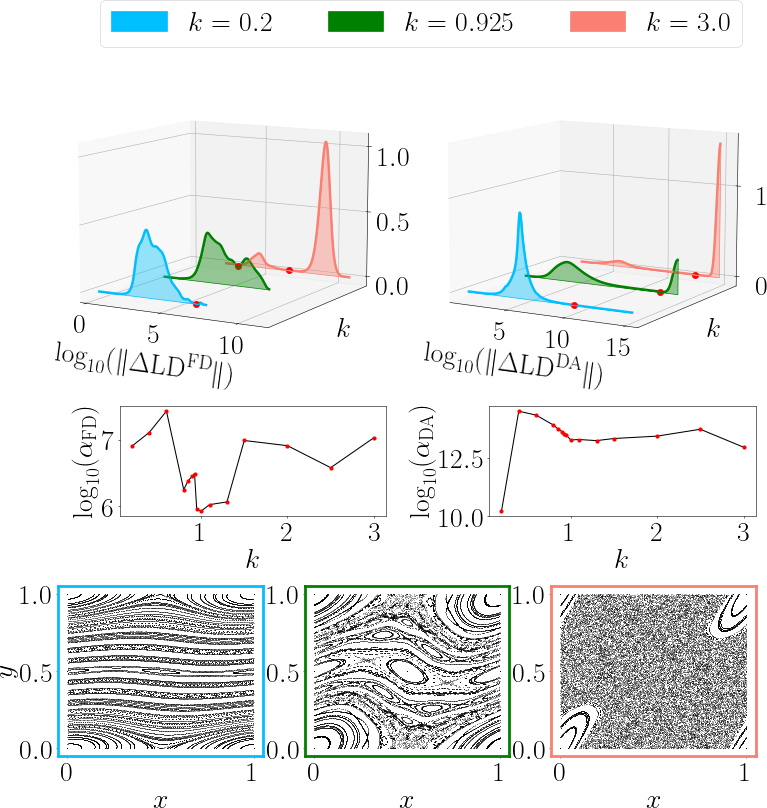}
    \caption{
    Composite plots showing representative distributions and their corresponding thresholds $\alpha$ (red bullet points)
    obtained with $\dldfd$ or $\dldda$ on the standard map with increasing values of the nonlinearity parameter $k$. 
    The distributions  are representative of phase spaces, shown in the bottom row, dominated by invariant structures, a mix of stable and chaotic orbits, and chaos, respectively. 
    The distributions obtained with 
    $\dldda$ are cleaner and sharper, facilitating the determination of the thresholds 
    $\alpha$ to be used  to classify binary ordered and chaotic trajectories.}
    \label{fig:Distributions2D}
\end{figure}

The composite Fig.\,\ref{fig:Distributions2D} presents distributions of $\dldfd$ and $\dldda$  computed on  $[0,1]^{2}$ at time $N=10^3$. DA iterations  are stopped whenever $\log_{10}(\dldda) \ge T=15$ for the first time or the number of iterations exceeds $N$.
The three $k$ values considered are representative of phase spaces dominated by invariant structures ($k=0.2$), a balanced mix of regular and chaotic orbits (mixed phase space, $k=0.925$),  and a phase space dominated by chaos ($k=3$). 
The different shapes of the $\dld$ distributions demonstrate the sensitivity of the index with respect to the numerical method employed to estimate the second derivatives (even with the fine scale $j=9$).
As previously claimed, we can see that the $\dldda$ profiles are sharper and cleaner compared to the 
$\dldfd$ distributions. This is  especially visible for the parameters $k=0.2$ and $k=0.925$, where the spikes in the $\dldda$ distributions are much more pricked, clear, and separated. 
On each distribution, the red bullet dot indicates the threshold 
$\alpha$ selected according to the distribution based method (see Sect.~\ref{sec:methods}) and used to classify orbits as regular or chaotic.   
The DA arithmetic leads to sharp differences in the values taken by $\dldda$ according to the dynamical regime, facilitating the determination of the thresholds $\alpha$, which are shown in the middle panels of Fig.~\ref{fig:Distributions2D} for $\dldfd$ and $\dldda$ methods.

\begin{figure}
    \centering
    \includegraphics[width=0.9\textwidth]{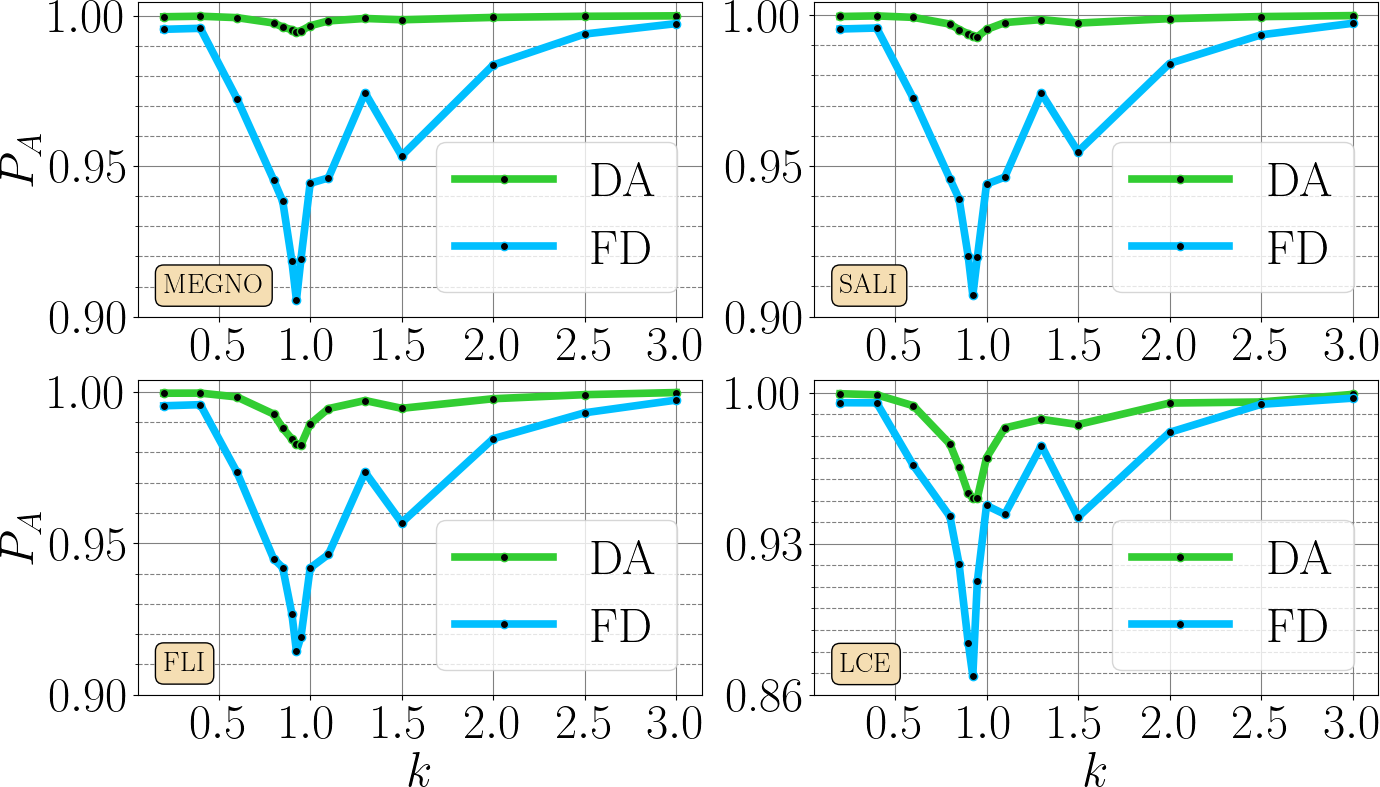}
    \caption{
    Performances $P_{A}$ of the $\dldfd$ and $\dldda$ indices at $N=10^{3}$ on the standard map  across
    a range of perturbation values $k$ against tangent map based methods. MEGNO, SALI, and FLI are computed at $N=10^{3}$, LCE is computed at $N=10^{7}$. $\dldda$ is nearly as sensitive as tangent map based methods across the whole range of perturbation parameter $k$.  
    }
    \label{fig:fig3}
\end{figure}

\begin{figure}
    \centering
    \includegraphics[width=0.8\textwidth]{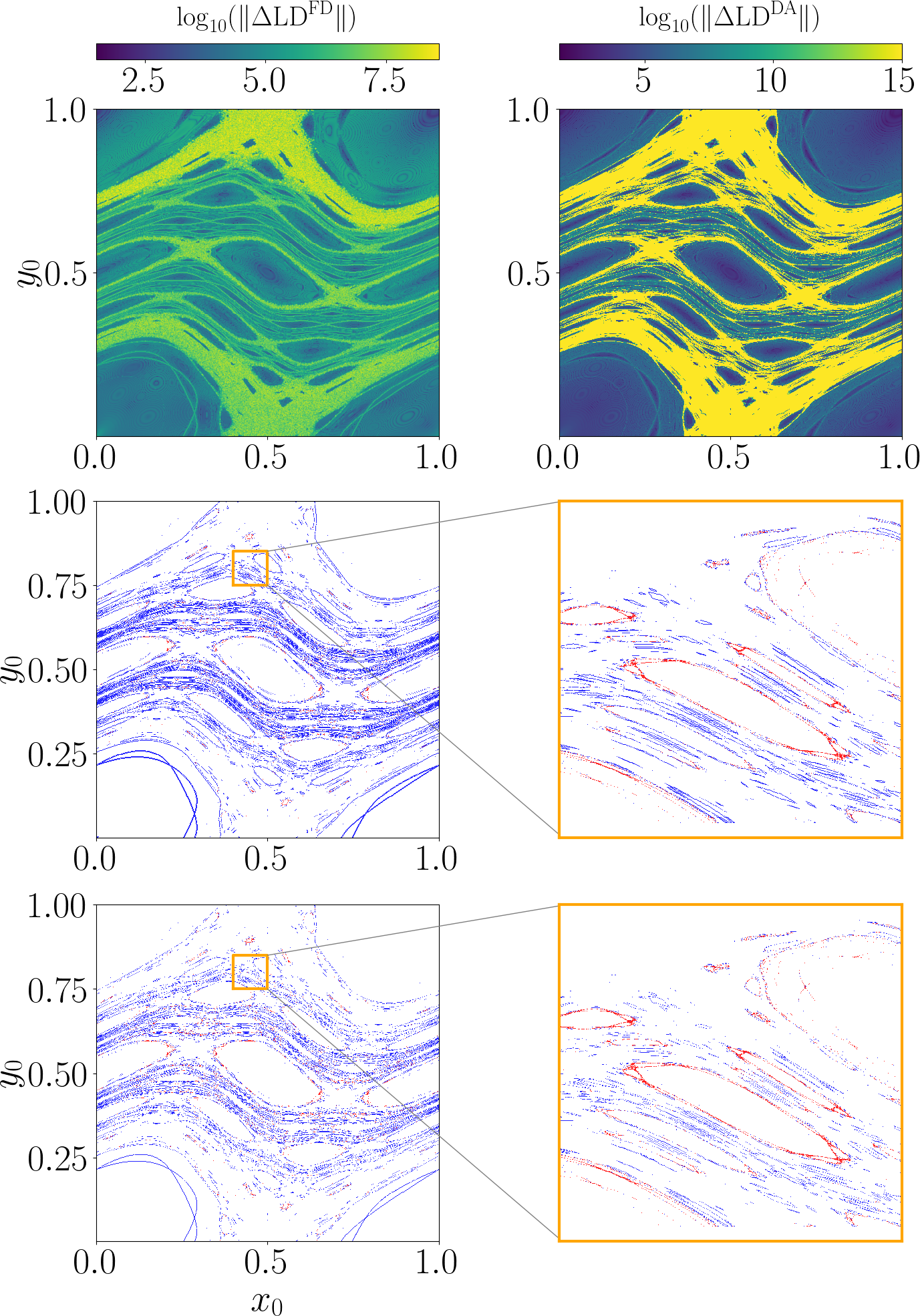}
    \caption{
    (Top row)
    $\log_{10}(\dldfd)$ and $\log_{10}(\dldda)$ maps for the standard map with $k=0.925$ computed at time $N=10^{3}$. The $\dldda$ map provides a sharper contrast of the various dynamical structures. 
    (Middle row) Two scale plot analysis of the long-term stability,  obtained with the mean MEGNO over $N=10^{5}$ iterations, 
    for the mismatch set $\mathcal{M}$ of $\dldfd$ and $\dldda$ (which are also computed for $10^5$ iterations). The blue color encodes mean MEGNO stable orbit, whilst the red color encodes mean MEGNO chaotic orbit. 
    Most initial conditions in the mismatch set $\mathcal{M}$ are  close to dynamical structures and are mean MEGNO stable over the timescale investigated. (Bottom row) Same as before, but the $\dldfd$ map is obtained with a meshed domain at order $10$ (instead of 9).  
    }
    \label{fig:fig2}
\end{figure}

In order to assess quantitatively the differences and sensitivity 
of $\dldfd$ and $\dldda$
in discriminating stable and chaotic trajectories, we compute a simple probability of agreement $P_{A}$ (also used in \cite{mHi22}), 
corresponding to the fraction of orbits that are correctly classified (using either methods) against the results provided with tangent map based methods\footnote{
The threshold values set for binary classification with the various tangent map methods are also determined by examining  their  distributions.} considered as ground truth (see their definitions provided in \ref{sec:Appendix2}).
For a set of initial conditions, it reads 
\begin{align}\label{eq:PA}
    P_{A}= 
    \frac{\# \textrm{Orbits correctly classified}}{\# \textrm{Number of initial conditions}}
    .
\end{align} 
Fig.\,\ref{fig:fig3} concatenates the probabilities $P_{A}$ of $90$  dynamical maps computed across a range of $15$ perturbation values $k$ spanning $\mathcal{I}=[0.2,3]$. 
The observables $\dldda$, $\dldfd$, MEGNO, FLI, and SALI are computed for $N=10^{3}$ iterations, while 
LCE, taking longer to converge,  is computed at $N=10^{7}$ iterations.  
The probability of agreement curves obtained against mean MEGNO, SALI and FLI are nearly identical but differ  slightly with the LCE results. 
A number of observations can be made in reference to Fig.\,\ref{fig:fig3}. 
First, the probabilities confirm  that $\dldfd$ shows overall good performances against the tangent map based methods tested as, in the worst case scenario, $P_{A}$ is above $87\%$ over the range of parameters $k$ investigated. It thus furnishes an cheap but efficient method to distinguish chaos. 
Second, the classification accuracy of 
$\dldda$  outperforms the $\dldfd$ index, pointing out the benefits of DA arithmetic and the finite-size effects of the meshed domain on $\dldfd$. 
In fact, the classification 
achieved with $\dldda$ is always and astonishingly above $98\%$ accuracy against mean MEGNO, SALI, and FLI, and very stable (in fact, almost constant) across the whole range of parameters $k$.
This contrasts with the $\dldfd$ results showing similar performances only either in the very stable ($k < 0.5$) or very chaotic regimes $(k > 2.5)$. 
For the perturbation parameters corresponding to a mixed phase space, \ie \, $k \sim 1$, $\dldfd$ shows a sharp decrease of its performance. 
For this dynamical regime and under our computational settings, the misclassifications against $\dldda$ might reach $\sim 10\%$.

\begin{figure}
\centering
\includegraphics[width=1\textwidth]{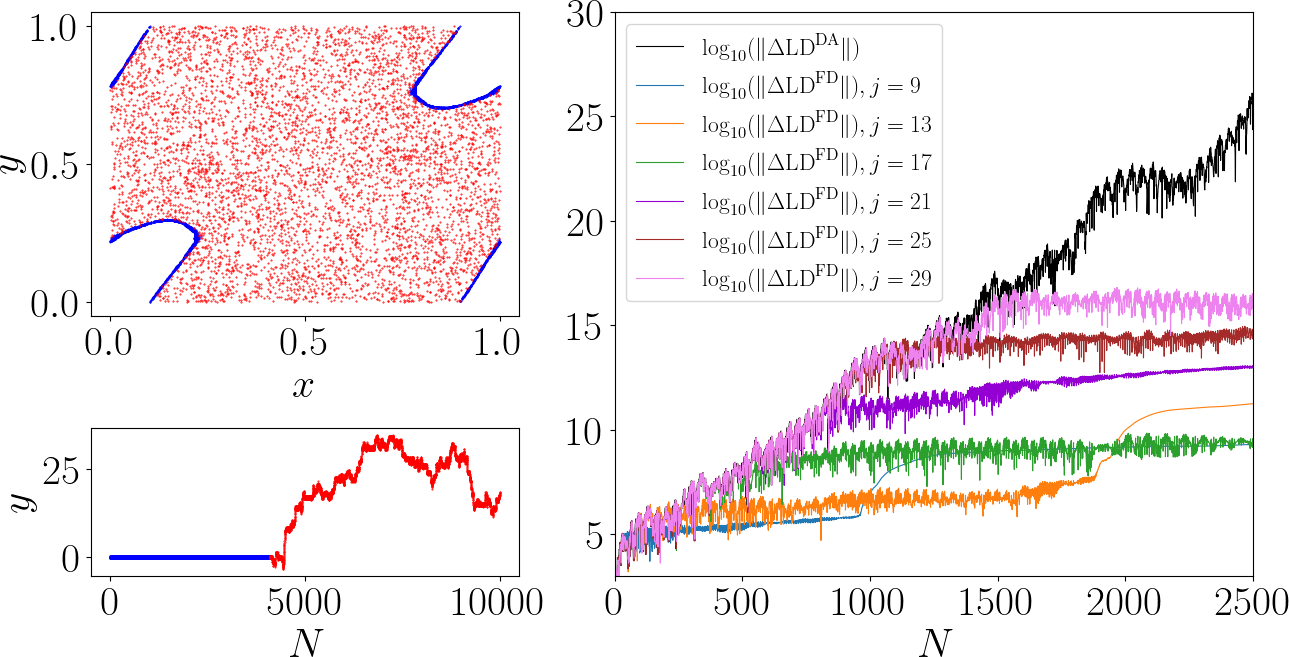}
\caption{
Composite plot showing the time evolution of a sticky orbit up to time $N=10^{4}$ in the phase space and the time evolution of the its $y$-component. 
The escape time is close to $N=4.1 \times 10^{3}$. Also shown are the time evolutions up to $N=2.5 \times 10^{3}$ of the $\dldda$ and $\dldfd$ indicators, the latter being computed for increasing values of $j$. The larger $j$, the closer $\dldfd$ to $\dldda$. The nonlinearity parameter is set to $k=3$ and the initial condition is $x_{0}=0.0088046875$ and $y_{0}=0.23123828125$.
}
\label{fig:2DSticky}
\end{figure}

For a more in-depth analysis of the misclassifications, the first row of Fig.\,\ref{fig:fig2} shows, for $k=0.925$ (the perturbation value leading to the largest discrepancy), the 
$\log_{10}(\dldfd)$ and $\log_{10}(\dldda)$ dynamical maps at time $N=10^{3}$. Although the maps look globally alike and both delineate the dynamical structures,  
$\dldda$ provides a sharper contrast in distinguishing  ordered and chaotic trajectories. Moreover, the thin Moiré-like patterns visible in the librational zones of the $\dldfd$ maps tend to be  less prominent in the $\dldda$ map. 
The second row of Fig.\,\ref{fig:fig2} shows a two-scale analysis (materialised by the orange box) of the initial conditions for which the $\dldfd$ and $\dldda$ classifications disagree (hereafter refereed as the mismatch set $\mathcal{M}$). 
For this choice of nonlinearity parameter, one notices that $\mathcal{M}$ contains  mostly initial conditions on the edges of thin dynamical structures along separatrices and chaotic bands near periodic chains of the phase space. This is best observed in the zoomed-in portion. 
We characterised further the long-term stability of points of $\mathcal{M}$ by computing the mean MEGNO at time $N=10^{5}$. 
Mean MEGNO points of $\mathcal{M}$ that are regular appear in blue, whilst chaotic mean MEGNO orbits are coloured in red. 
It appears that most of the points in $\mathcal{M}$ are mean MEGNO stable 
($96\%$ mean MEGNO stable at the global scale, $75\%$ mean MEGNO stable at the smaller scale). 
We repeated the former computation but this time estimating $\dldfd$ on a grid at order $10$ (and still comparing with $\dldda$ on the $j=9$ grid) in the third row of Fig.\,\ref{fig:fig2}. The disagreement between the $\dldfd$ and $\dldda$ classification passed from $11.5 \%$ to $7.6\%$ at the global scale and from $4.6 \%$ to $3.5 \%$ at the smaller scale. 
Those results suggest that working with finite differences, one should increase the parameter $j$ of the mesh to at least $j=11$ to compete with the results of $\dldda$ computed on a $j=9$ grid. 
To illustrate further this aspect, we selected the initial condition of a sticky orbit of the mismatch set $\mathcal{M}$ for $k=3$.
Its time evolution in the phase space and the time evolution of the $y$ coordinate are shown in Fig.\,\ref{fig:2DSticky} up to time $N=10^{4}$. 
We colored segments of the orbit with different colors before and after the time $N=4.1 \times 10^{3}$, corresponding approximately to the time spent near the main stable island (blue color), before the orbit's escape within the surrounding chaotic sea (red color, scattered region). 
For this particular initial condition, $\dldfd$ computed with the default order $j=9$ fails to recognise the chaoticity of the orbit at time $N=10^{3}$, while $\dldda$ correctly identifies its chaotic nature, as shown in the right panel of Fig.\,\ref{fig:2DSticky}. Actually, the thresholds determined from the distributions are $\alpha=7.03$ for $\dldfd$ and $\alpha=12.94$ for $\dldda$. Fig.\,\ref{fig:2DSticky} also shows the time evolution of $\dldfd$ computed for increasing orders $j$.
Albeit we are unable to give a ruling regarding the mismatch set for these orders, we can appreciate that the higher $j$, the closer $\dldfd$ to $\dldda$.
Up to time $N=10^{3}$, $\dldfd$ recovers $\dldda$ at order $j=25$.

Finally, the reliability of our results with respect to the choice of the thresholds for $\dldfd$ and $\dldda$ are further examined in \ref{sec:Appendix3}. We show that small variations in the thresholds strongly affect the performance of $\dldfd$ in discriminating stable and chaotic trajectories, especially in mixed phase space, while they only slightly affect the stability predictions given by $\dldda$ thanks to its clearer and sharper distribution.

%=========
\subsection{Applications to $2$ coupled standard maps} \label{subsec:4d}
%=========
This section repeats the steps of \autoref{sec:resultsSM} on the following $4$ dimensional mapping 
\begin{align}\label{eq:StandardMap4D}
\left\{
    \begin{aligned}
    &x'_1=x_1+y'_1, \\
    &x'_2=x_2+y'_1, \\
    &y'_1=y_1-\frac{1}{2 \pi } \big(a \sin(2 \pi x_1)+c\sin(2\pi(x_1+x_2))\big),\\
    &y'_2=y_2-\frac{1}{2 \pi } \big(b \sin(2 \pi x_2)+c\sin(2\pi(x_1+x_2))\big),
    \end{aligned}
\right.
\end{align} 
where $a$, $b$, $c$ represent parameters.
The case $c=0$  reduces the dynamics to a product of $2$ uncoupled standard maps. 
We set $a=b=0.2$ and consider  $c$ as a free parameter  varying in $\mathcal{I}=[0.01,2.3]$. The  $4$ dimensional dynamics is visualised over $2$ dimensional sections, by freezing 
$x_1=0$ and $x_2=0$.  The remaining initial conditions $(y_{1},y_{2})$ evolve in 
$D=[-0.25,0.65]^2$. \\

\begin{figure}
    \centering
    \includegraphics[width=\textwidth]{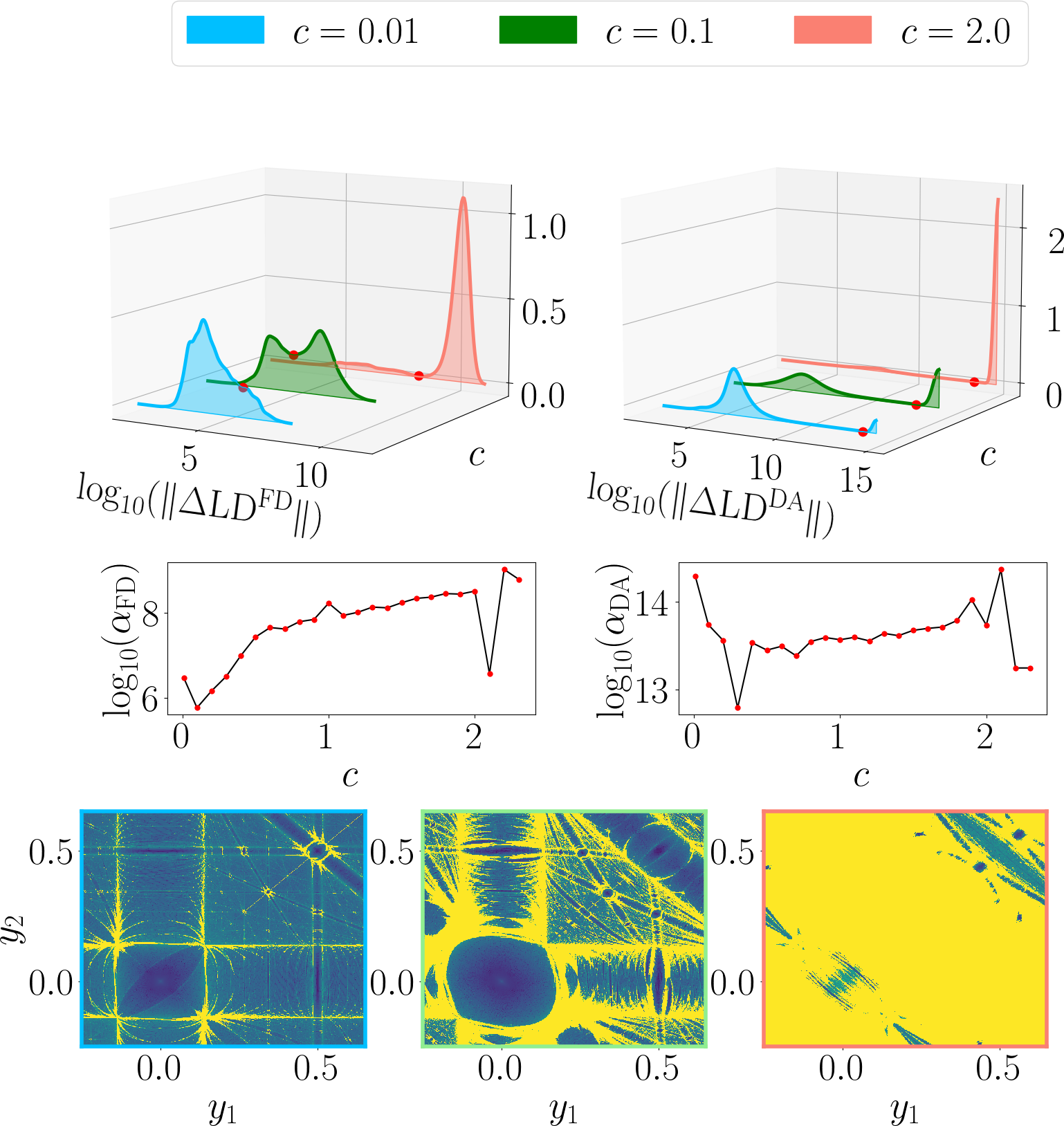}
    \caption{
    Composite plots showing representative distributions and their corresponding thresholds $\alpha$ 
    obtained with $\dldfd$ or $\dldda$ on the four-dimensional mapping of Eq.\,(\ref{eq:StandardMap4D}) with increasing values of the coupling parameter $c$ in the $(y_{1},y_{2})$ plan. 
    The distributions  are representative of phase spaces, shown in the bottom row and computed with $\dldda$, dominated by invariant structures ($c=0.01$), mixed phase space ($c=0.1$), and chaos ($c=2$), respectively. 
    The distributions obtained with 
    $\dldda$ are cleaner and sharper, and greatly ease the determination of the thresholds 
    $\alpha$ to be used with $\dld$ in order to classify ordered and chaotic trajectories. This is especially visible by comparing the distributions at low value for $c=0.01$, where the $\dldfd$ distribution is spread-out without clear spikes.}
    \label{fig:Distributions4D}
\end{figure}

Figure \ref{fig:Distributions4D} is the analogue of Fig.\,\ref{fig:Distributions2D} for Eq.\,(\ref{eq:StandardMap4D}) and shows the distributions of the $\dldfd$ and $\dldda$ indices together with the  thresholds $\alpha$ derived from them for three distinct dynamical regimes (dominated by regular orbits, mixed phase space, and chaos). The bottom row displays $\dldda$ dynamical maps. 
As previously noticed, the distributions obtained with $\dldda$ are cleaner. 
In particular, for the smallest considered perturbation value, $c=0.01$, the distribution of $\dldda$ still contains two clearly identifiable peaks, contrarily to the $\dldfd$ distribution. The spike at large $\dldda$ values correspond to the large values taken by $\dldda$ on and close to separatrice crossings, where located  chaos is found to exist. For this particular value, $\dldfd$ is not really bimodal, but is rather fat and spread-out. In particular, the distribution does not chiefly reflect the existence of thin chaotic layers. In this case, the threshold $\alpha$ determined from the $\dldfd$ distribution is certainly questionable. \\

\begin{figure}
    \centering
    \includegraphics[width=0.7\textwidth]{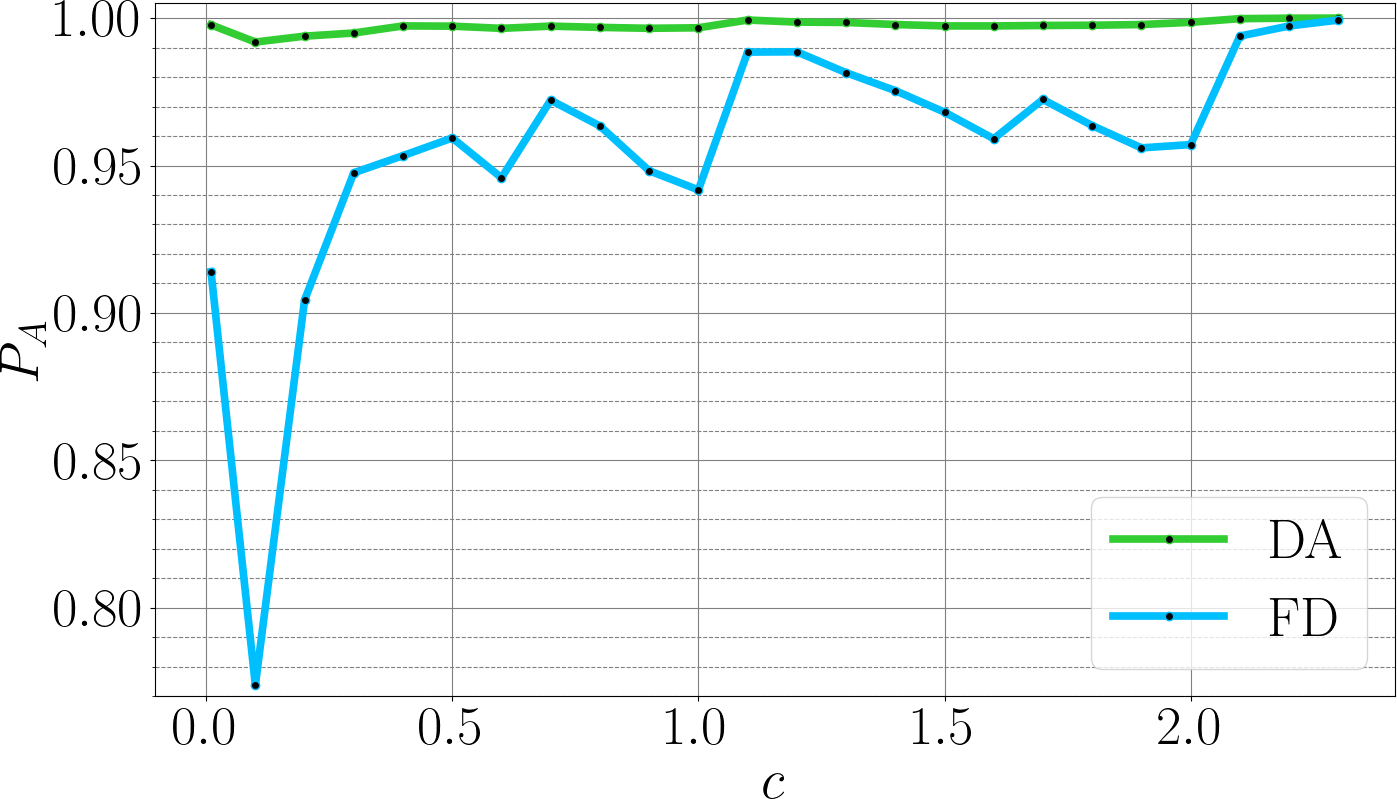}
    \caption{
    Performances $P_{A}$ of the $\dldfd$ and $\dldda$ indices on the 4-dimensional model of Eq.\,(\ref{eq:StandardMap4D})
    across a range of coupling values $c$ against the mean MEGNO indicator at $N=4 \times 10^{3}$. Differential arithmetic leads to very stable performances.}
\label{fig:MismatchEvolution4DStdMapMegno}
\end{figure}

\begin{figure}
    \centering
    \includegraphics[width=0.8\textwidth]{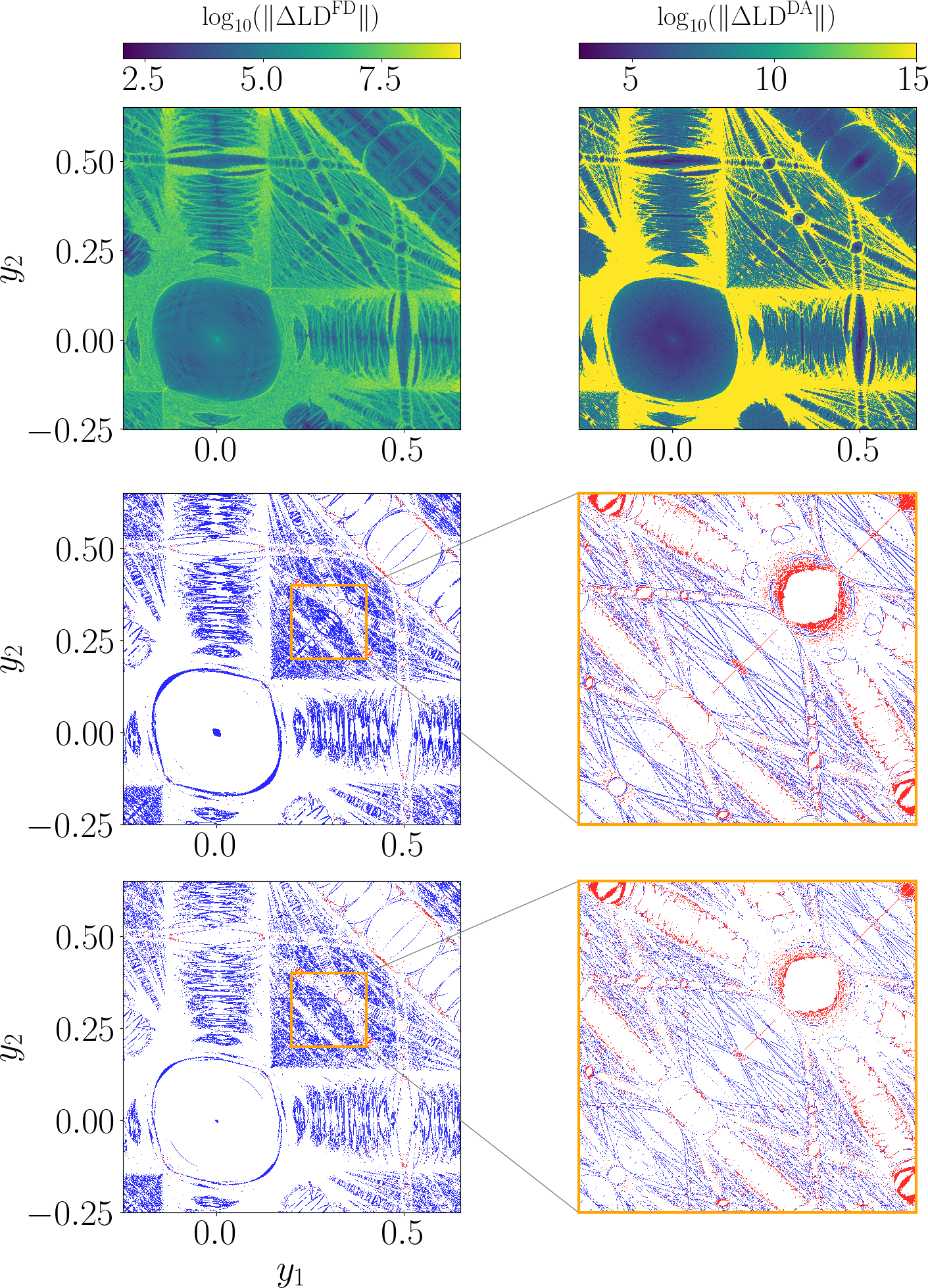}
    \caption{
 (Top row)
$\log_{10}(\dldfd)$ and $\log_{10}(\dldda)$ maps for Eq.\,(\ref{eq:StandardMap4D}) with $c=0.1$ computed at time $N=4 \times 10^{3}$. 
The $\dldda$ map provides a sharper contrast of the various dynamical structures. 
(Middle row) Two scale plot analysis of the long-term stability,  obtained with the mean MEGNO over $N=10^{5}$ iterations, 
for the mismatch set between $\mathcal{M}$ of $\dldfd$ and $\dldda$. The blue color encodes mean MEGNO stable orbit, whilst the red color encodes mean MEGNO chaotic orbit. 
Most initial conditions in the mismatch set $\mathcal{M}$ are  close to dynamical structures and are mean MEGNO stable over the timescale investigated.   
(Bottom row) Same as before, but the mismatch set is computed from $\dldfd$ on a grid with order $j=10$. Given the similarity between the middle and bottom rows, we infer that at least $j=11$ would be necessary to reduce the size of the mismatch set $\mathcal{M}$ for this dynamical regime, which is computationally costly.   
    }
\label{fig:4DPhasePortrat}
\end{figure}

Fig.\,\ref{fig:MismatchEvolution4DStdMapMegno} shows the probabilities $P_{A}$ of Eq.\,(\ref{eq:PA}) obtained from dynamical maps compared to the mean MEGNO classification at times 
$N=4 \times 10^{3}$ for a range of coupling values of $c$. 
Over the whole range of values, $\dldda$ outperforms $\dldfd$. 
As for the 2 dimensional case, the $\dldda$ performances are remarkably stable and as sensitive as the mean MEGNO, being $P_{A}$ larger than $99\%$. 
The $\dldfd$ index shows comparable performances ($P_{A} \ge 95\%$) only in the very chaotic domain, for large $c$ values ($c > 1$). 
In the weak coupling cases, for $c<0.2$, the performances of $\dldfd$ sharply decrease and  
the maximal difference between $\dldfd$ and $\dldda$ reaches $22\%$ for $c=0.1$. 
For the latter value, the top row of Fig.\,\ref{fig:4DPhasePortrat} shows respectively 
the $\dldfd$ and $\dldda$ maps computed at $N=4 \times 10^{3}$, together with a two-scale analysis of the long-term stability, obtained with mean MEGNO at $N=10^{5}$, of the mismatch set $\mathcal{M}$. 
We can see that this perturbation value corresponds to a phase space where many of the dynamical structures project as thin chaotic lines around high-order resonances. Overall,
the $\dldda$ index provides better visual contrast of the dynamical structures appearing in the $(y_{1},y_{2})$ sections compared to $\dldfd$.
At the global scale $D=[-0.25,0.65]^{2}$, we found $96\%$ of the initial conditions of the mismatch set 
$\mathcal{M}$ to be stable. 
At the smaller scale we investigated, $D'=[0.2,0.4]^{2}$, we found that $61.63 \%$ of initial conditions of the mismatch set are mean MEGNO stable. 
The small scale analysis highlights very distinctively the
distribution of the points of $\mathcal{M}$
along the edges and thin resonances of the system. 
This is a consequence of finite differencing neighboring points, which tend to approximate and smooth out the locations of the hyperbolic structures (see \eg \, \cite{sSh05} in the context of Lagrangian Coherent Structures detection). Finally, the last row of Fig.\,\ref{fig:4DPhasePortrat} shows the mismatch set obtained by computing $\dldfd$ on a grid order $j=10$ with the results given by $\dldda$ computed for $j=9$. The extreme resemblance of the second and third rows, together with extremely similar statistics, demonstrate that $j=10$ would still not be enough to discriminate accurately chaotic from regular trajectories based on finite differences. This suggests that for this dynamical regime, at least $j=11$ would be required, increasing significantly the computational burden of finite differences methods.

\section{Applications to non-uniformly sampled meshes of initial conditions}\label{sec:NU}
To challenge further FD methods, we focus in this section on non-uniformly sampled LDs fields for the $2$ dimensional standard map  on $D=[0,1]^{2}$ for $k=0.925$ and $N=10^3$ iterations.  
The non-uniform LD fields are generated by dowsampling,  according to a uniform distribution, up to a maximal percentage (specified below) LD values from the initial uniform fields.  
From the downsampled non-uniform LD fields, one still seeks to estimate $\dldfd$ and compare their performances against $\dldda$. \\

\begin{figure}
    \centering
    \includegraphics{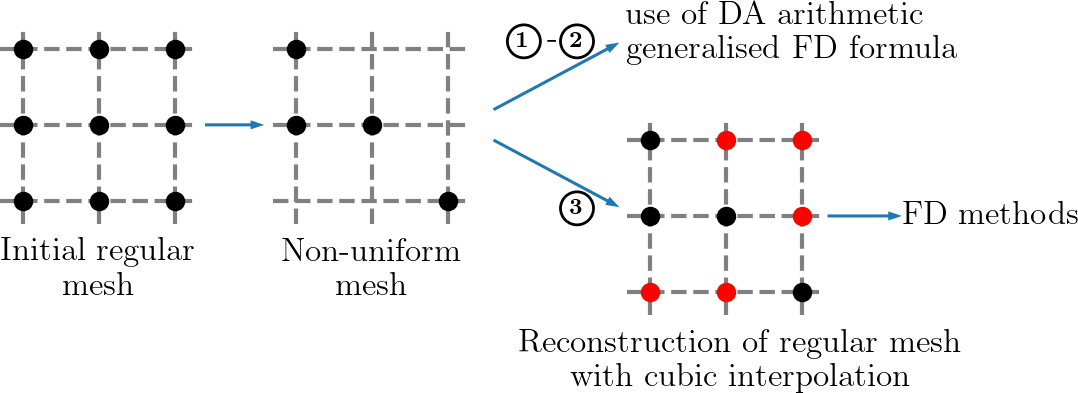}
    \caption{
Sketch illustrating how irregular meshes are generated and managed. 
From a uniformly sampled $2$-dimensional LD field, we generate a non-uniform LD field by removing a certain percentage of nodes according to a uniform distribution. 
In order to estimate $\dld$ from this field, we follow three strategies: 
\ding{172} Thanks to differential arithmetic, we have access to $\dldda$ that approximates $\dld$ up to machine precision directly from the non-uniformly sampled field,
\ding{173} We estimate $\dldfd$ by using generalised-finite differences formulas on the non-uniformly sampled mesh - see Eq.\,(\ref{eq:DerivativesForNonUniformGrid}), 
and 
\ding{174} We use a cubic interpolation to reconstruct a uniformly sampled LD field, from which usual finite difference formulas of Eq.\,(\ref{eq:DLDFD}) apply.
}
\label{fig:NonUniformFlowChar}
\end{figure}

Under this setting, the DA framework appears to be beneficial given that the derivatives are already known, also for the nodes of the non-uniform mesh. On the contrary, $\dldfd$ requires the computation of LD values at neighboring points which are not in the non-uniform mesh. 
To estimate $\dldfd$ from the non-uniform LD field,  we follow two strategies:
\begin{enumerate}
    \item We use a generalised  central difference formula taking into account the non-uniform sampling. For illustrative purpose, let us consider $m+1$ values of a real observable $f$, 
    $\{f_{0},f_{1},\cdots,f_{m}\}$, at coordinate points 
    $\{x_{0},x_{1},\cdots,x_{m}\}$ ordered increasingly. We define 
    $h_{i}=x_{i+1}-x_{i}$, $i=0,\dots,m-1$. The following approximations hold (see \cite{jLe18}):
\begin{align}\label{eq:DerivativesForNonUniformGrid}
    f^{'}_{i} \simeq
    \left\{
    \begin{aligned}
   & \frac{h_{i-1}^2 f_{i+1}-(h_{i-1}^2-h_i^2)f_i-h_i^2 f_{i-1}}{h_{i-1}h_i(h_{i-1}+h_i)}, \, \forall i \in \{ 1,\dots,m-1 \}, \\
   & \frac{-h_{0}^2 f_{2}+(h_0+h_1)^2 f_1+(h_0^2-(h_1+h_2)^2) f_{0}}{h_0h_1(h_{0}+h_1)}, i=0, \\
   & \frac{h_{m-1}^2 f_{m-2}-(h_{m-2}+h_{m-1})^2 f_{m-1}-(h_{m-1}^2-(h_{m-1}+h_{m-2})^2) f_{m}}{h_{m-2}h_{m-1}(h_{m-2}+h_{m-1})}, \, i=m.
   \end{aligned}
    \right.
\end{align}
In order to estimate $\dldfd$ on non-uniform grids with finite differences, we iterate the above formulas to obtain the second order derivative
with $f=\textrm{LD}$ and repeat the process along each direction of the mesh. 
    \item We reconstruct a uniform LD field using cubic interpolation of the LDs. Once the uniform LD field is reconstructed, we apply the standard FD formula given in Eq.\,(\ref{eq:DLDFD}). The interpolated uniform field is computed with the \texttt{SciPy} library and the
    \texttt{interpolate.griddata} function \cite{pVi}.
\end{enumerate}
This workflow is summarised in Fig.\,\ref{fig:NonUniformFlowChar}. \\

\begin{figure}
    \centering
    \includegraphics[width=0.7\textwidth]{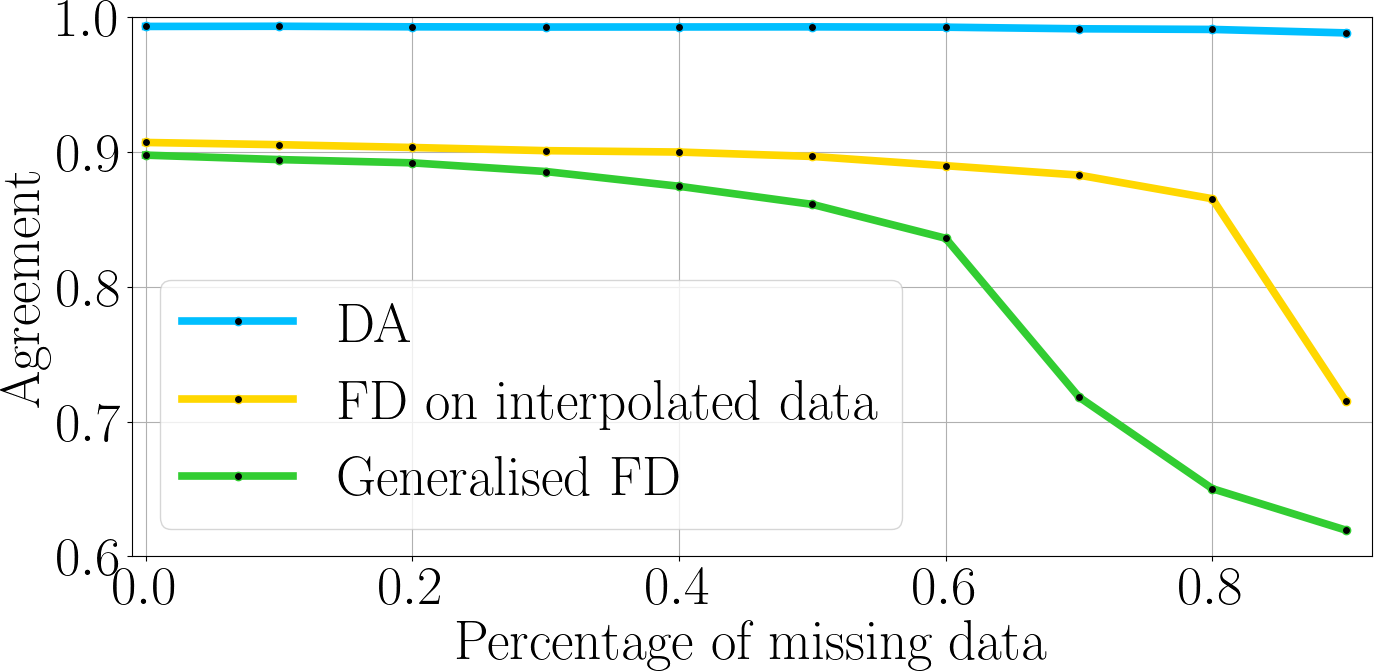}
    \caption{Performances of $\dldda$ and $\dldfd$ on non-uniform LD field.}
\label{fig:2DNonUniform}
\end{figure}

\autoref{fig:2DNonUniform} shows the performances $P_{A}$ as a function of the percentage of deleted points.  The performance of the $\dldda$ indicator is constant with $P_{A}$ very close to $100\%$, while the others two methods, starting at $P_{A}=90\%$, decrease slowly in performance. 
The FD formula performed on the interpolated field has comparable performances with generalised FD formula up to $60\%$ of the initial points removed. At this value, the two methods start to separate more distinctively. 
Working with the interpolated field provides  better results compared to the generalised FD formula of Eq.\,\eqref{eq:DerivativesForNonUniformGrid}. 
This is true even when there are not missing points, in which case formula \eqref{eq:DerivativesForNonUniformGrid} becomes
\begin{equation} \label{eq:NonUniformInUniformCase}
   4 h^2 f_i''=f_{i+2}-2f_i+f_{i-2}
\end{equation}
and the interpolation method reduces to use of \eqref{eq:DLDFD}.

\section{Conclusions} \label{sec:conclusions}
This paper further complemented the theory of Lagrangian descriptors for chaos detection based on the arc length of trajectories.
The main contributions and conclusions of this study are the following.
\begin{enumerate}
    \item The $\dld$ index  proposed as fast chaos indicator in \cite{jDa22} is based on 
    the second derivatives of the arc-length metric with respect to initial conditions, computed over a finite time window.  
    $\dld$ was so far approximated by finite differencing techniques on fine meshed domain via $\dldfd$, as in Eq.\,(\ref{eq:DLDFD}). 
    The cornerstone of this contribution is the proposal of differential algebra, as a mean of automatic differentiation, to introduce the $\dldda$ index. 
    Contrarily to $\dldfd$, 
    $\dldda$ is free of mesh properties and approximate $\dld$ up to machine precision.  
    \item Based on the $2$-dimensional standard map model and higher $4$-dimensional coupled versions, we performed a  comprehensive  quantitative performance assessment of the  $\dldfd$ and $\dldda$ indices. 
    We confronted the latter to a set of well established tangent map based chaos indicators, including  the mean MEGNO, the fast Lyapunov indicator, the smaller alignement index and the finite-time Lyapunov characteristic exponent. We explored with care parametric dependencies of the models to compute global statistics and the sensitivity of the indicators. We summarise our results as follows.
    \begin{enumerate}
        \item We confirmed that $\dldfd$ is a viable fast indicator to portray phase spaces, with performances over $90\%$ for the standard map. 
        For the 4-dimensional model, the performances degrade in the mixed phase space regime and we noted that up to $20\%$ of orbits might be misclassified. Under this setting, care must be taken if, besides global visualisation,  discrimination of orbits is needed with $\dldfd$. 
        \item Having computed and compared systematically performances of $\dldfd$ and $\dldda$, we provided evidences that mesh properties influence the discrimination of orbits  when using $\dldfd$.  For typical mesh resolution  ($j=9$) and for the timescale investigated, errors in the classification between $\dldfd$ and $\dldfd$ can reach at worst $10\%$ for the standard map (see Fig.\,\ref{fig:fig2}) and up to $22\%$ in the nearly integrable $4$-dimensional mapping (see Fig.\,\ref{fig:4DPhasePortrat}).  
        \item Most of the misclassified initial conditions with $\dldfd$ and $\dldda$ (mismatch set) appear to be located on the edges of the dynamical structures. Over the timescale investigated, the majority of them are classified as regular with the mean MEGNO indicator.
        \item Contrarily to $\dldfd$, global statistics of $\dldda$ show an astonishingly good agreement with fast tangent based chaos indicators (FLI, mean MEGNO, SALI), over $98\%$ of agreement in general. Thus, $\dldda$ might be considered as sensitive as fast indicators based on tangent map dynamics. 
        \item The threshold values $\alpha$ to be used with $\dld$ for the binary classification of initial conditions (see Eq.\,(\ref{eq:ThresholdsDelta})) were discussed in the context of phase spaces dominated by stability, chaoticity, or a balanced mix of both. 
        The DA framework leads to clear unimodal or bimodal distributions, contrarily to finite differences that might lead to fat-tailed distributions without clearly identifiable peaks, for which the recommended procedure for determining the threshold is certainly not optimal. 
        Thus, DA arithmetic leads to a much more robust threshold determination procedure.  
    \end{enumerate}
    \item The benefits of the DA framework were further demonstrated by considering  the performances of the $\dldfd$
    and $\dldda$
    indices on non-uniform depleted meshes of initial conditions, a situation that   challenges finite differencing approaches. 
\end{enumerate}
In summary, $\dld$ is a robust and sensitive chaos indicator to portray easily the qualitative geometry of the phase space. 
For precise classification of initial conditions, the use of DA arithmetic and the associated $\dldda$ indicator is preferable. Statistically, this indicator appears to be as sensitive as tangent map methods.  
Extension of our observations and conclusions to flows, as their scaling with the dimension of dynamical systems, calls for future efforts. 

\section*{Acknowledgments}
The work of A. C\u{a}liman and A.-S. Libert is part of the Concerted Research Actions financed by the F\'ed\'eration Wallonie-Bruxelles (CAML project). The authors acknowledge discussions and feedback from T. Carletti.

\appendix
%==============
\section{First-order tangent map based chaos indicators}
\label{sec:Appendix2}
We provide the definitions of the finite-time tangent map based indicators used in this work.

Let us consider the smooth mapping $x_{n+1}=M(x_{n})$, $n \in \mathbb{N}$, and $x_{0} \in \mathbb{R}^{d}$. Let $v_{0}$ denotes a unitary tangent (deviation) vector. 
The tangent map dynamics is given by
\begin{align}
\left\{
    \begin{aligned}
    &x_{n+1}=M(x_{n}), \\
    &v_{n+1}=DM(x_{n})v_{n}.  
    \end{aligned}
\right.
\end{align} 
At iterate $N$, the norm of the tangent vector satisfies
\begin{align}
    \norm{v_{N}}
    =
    \norm{\prod_{i=0}^{N-1}DM(x_{i})v_{0}}.
\end{align}
The indicators defined below are based on the time evolution of this norm. 
%===============
\subsection*{The Maximal Lyapunov Exponent}
%===============
The (finite) maximal Lyapunov Characteristic Exponent (mLCE) is defined by
\begin{align}
\textrm{mLCE}(N;x_{0},v_{0}) =
\frac{1}{N} \log_{10}
\norm{v_{N}}.
\end{align}
Asymptotically with large $N$ ($N \to +\infty$), $\textrm{mLCE} \to 0$ for regular orbits, and $\textrm{mLCE} \to \delta > 0$ for chaotic orbits \cite{cSk09}. 

%===========
\subsection*{The Fast-Lyapunov Indicator}
%=========== 
The Fast Lyapunov Indicator introduced in \cite{cFr97} does not include the time average of the mLCE to speed-up the discrimination between regular and chaotic orbits. Several definitions of the FLI  exist in the literature. Here, we use the following definition of the FLI (see \cite{eLe01} for a recent review of the FLI method):
\begin{align}
\textrm{FLI}(N;x_{0},v_{0}) = 
\sup_{n \le N} \log_{10}
\norm{v_{n}}.
\end{align}
For regular orbits, FLI grows as $\alpha_{\textrm{FLI}}(N)=\log_{10}N$ and linearly with $N$ for chaotic orbits. 

%==========
\subsection*{The mean-MEGNO indicator}
%==========
The mean-MEGNO indicator follows from the MEGNO indicator introduced in \cite{pCi03}. 
The MEGNO indicator is defined as
\begin{equation}
    Y(N;x_{0},v_{0})=
    \frac{2}{N} \sum\limits_{i=1}^{N} \log_{10} \Bigg( \frac{\|v_i\|}{\|v_{i-1}\|} \Bigg) i.
\end{equation}
The mean-MEGNO indicator is the time average of the last quantity
 \begin{align}
    \overline{Y}(N;x_{0},v_{0})=
    \frac{1}{N} \sum\limits_{n=1}^{N}Y(n;x_{0},v_{0}).
\end{align}
The asymptotic evolution of $\overline{Y}$ is as follows. $\overline{Y} \to 0$ for stable periodic orbits, $\overline{Y} \to 2$
for quasi-periodic orbits and orbits close to stable periodic orbits, and $\overline{Y} \propto N$ for chaotic orbits.

%===========
\subsection*{The SALI}
%===========
The smaller alignment index (SALI) introduced in \cite{cSk01} requires to follow the iterations of two distinct unitary deviations vectors $v_{0}^{(1)}$ and $v_{0}^{(2)}$.   
The SALI at time $N$ reads 
\begin{align}
\textrm{SALI}(N;x_{0},v_{0}^{(1)},v_{0}^{(2)})
=
\min \Bigg\{  
\Bigg\| 
\frac{v_N^{(1)}}{\| v_N^{(1)} \|}+\frac{v_N^{(2)}}{\| v_N^{(2)} \|}  
\Bigg\|, \Bigg\| 
\frac{v_N^{(1)}}{\| v_N^{(1)} \|}-\frac{v_N^{(2)}}{\| v_N^{(2)} \|}  
\Bigg\| \Bigg\}.
\end{align}
$\textrm{SALI} \to \delta > 0$ with $N$ for regular orbits and $\textrm{SALI} \to 0$ with $
N$ for chaotic orbits. 

%================
\section{Reliability of $\dldfd$ and $\dldda$ with respect to their thresholds}\label{sec:Appendix3}
\begin{figure}
    \centering
    \includegraphics[width=0.8\linewidth]{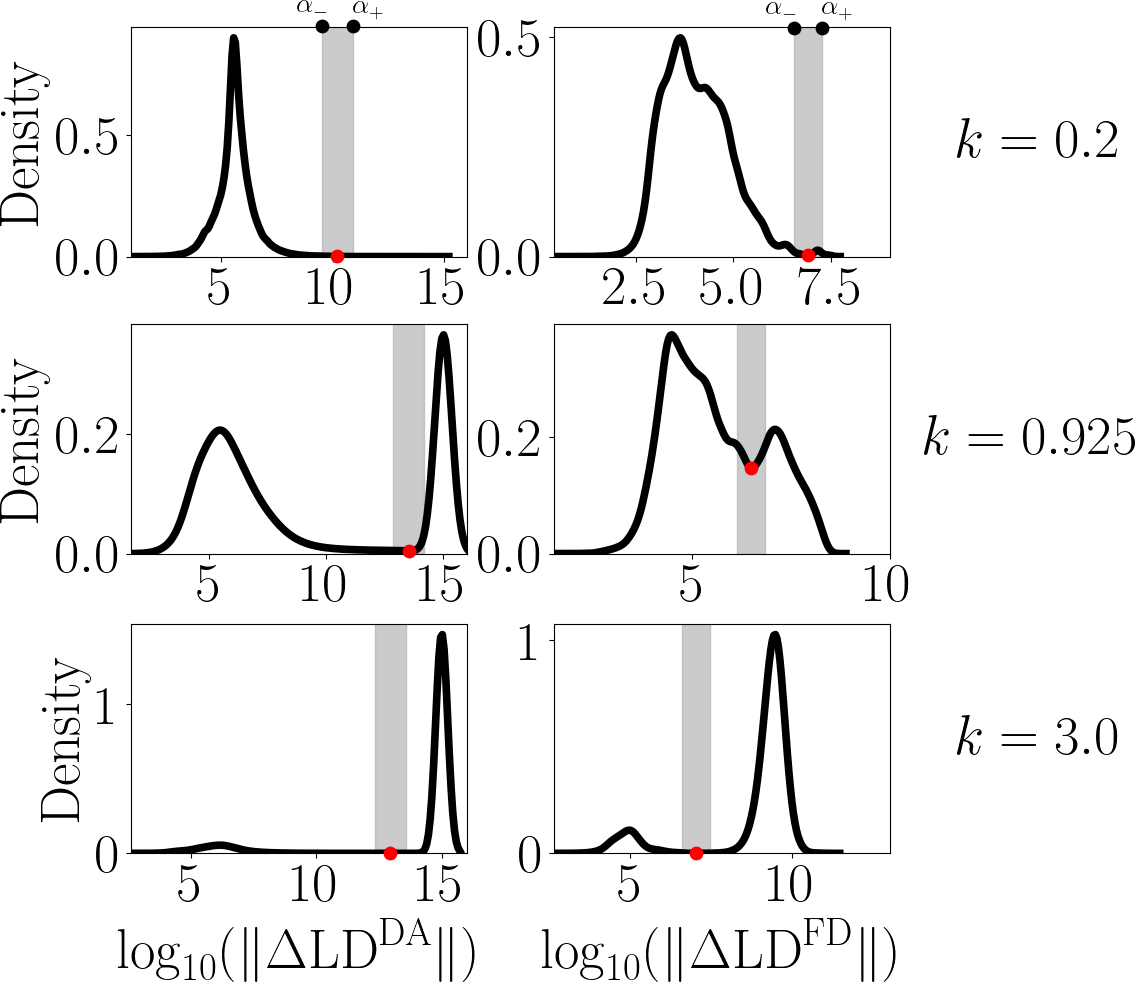}
    \caption{Small variations of the thresholds for $\dldda$ and $\dldfd$ used in our sensitivity analysis (see Table~\ref{table:DeltaLDThresholdsTable}).}
    \label{fig:DeltaLDThresholdsPicture}
\end{figure}

%==
In \autoref{subsec:dld}, we  explained how to select the threshold $\alpha$ for $\dldfd$ and $\dldda$ in order to classify in a binary way regular and chaotic trajectories. As previously described, the threshold $\alpha$ is set based on the shape of $\dldfd$ and $\dldda$ distributions obtained for many initial conditions. Fig.\,\ref{fig:DeltaLDThresholdsPicture} and Table~\ref{table:DeltaLDThresholdsTable} provide evidence that the results presented in Fig.\,\ref{fig:fig3} are sound and robust with respect to small variations of the chosen threshold. 
We conducted a sensitivity analysis by repeating our computations of the probability of agreement $P_{A}$, following the numerical setting of \autoref{sec:resultsSM}, for varying $\alpha$. 
We considered the $2$-dimensional standard map, iterated over $N=10^{3}$, and considered the same three values of the perturbation parameter $k$ that are representative of a phase space dominated by regular orbits, a mixed phase-space regime, and a phase space dominated by chaos. The results report the probabilities $P_{A}$ for $\alpha_{\pm} = \alpha \pm 0.05  \delta$, where $\delta$ is the maximal range of the $\dld$ values. The extremal $\alpha$ values considered are represented by a grey band in the distributions of Fig.\,\ref{fig:fig3}. The percentages of agreement associated to the probability $P_{A}$ are shown in Table~\ref{table:DeltaLDThresholdsTable} for the chosen $\alpha$ value and the two extremal ones. It is clear that the percentages vary very little for the $\dldda$ method, while the stability classification obtained with the $\dldfd$ method is highly sensitive on the choice of the threshold.

\begin{table} 
\centering
\begin{tabular}{c}
\begin{tabular}{c  c  c  c  } 
\vspace{-0.2cm}
\\
\multicolumn{4}{c}{$k=0.2$} \\
 \hline \hline \vspace{-0.3cm}
 \multirow{3}{*}{$\overline{Y}$}  &  &  &      \\
  & $\alpha_{-}$ & $99.88\%$ & $97.02\%$ \\ 
  & $\boldsymbol{\alpha}$ & $\boldsymbol{99.95\%}$ & $\boldsymbol{98.38 \%}$\\
 & $\alpha_{+}$ & $9.98\%$ & $99.29\%$ \\
 \hline \vspace{-0.3cm}
 \multirow{3}{*}{SALI}  &  &  & \\
  & $\alpha_{-}$ & $99.88 \%$ & $97.02 \%$ \\ 
  & $\boldsymbol{\alpha}$ & $\boldsymbol{99.95\%}$ & $\boldsymbol{98.38\%}$\\
 & $\alpha_{+}$ & $99.98 \%$ & $99.3 \%$ \\
 \hline  \vspace{-0.3cm}
 \multirow{3}{*}{FLI}  &  &  &     \\
  & $\alpha_{-}$ & $99.88\%$ & $97.02\%$\\ 
  & $\boldsymbol{\alpha}$ & $\boldsymbol{99.95\%}$ & $\boldsymbol{98.38\%}$ \\
 & $\alpha_{+}$ & $99.98\%$ & $99.3\%$ \\
 \hline\hline

 \multicolumn{4}{c}{$k=0.925$} \\
 
 \hline \hline \vspace{-0.3cm}
 \multirow{3}{*}{$\overline{Y}$}  &  &  &      \\
  & $\alpha_{-}$ & $99.26\%$ & $91.52\%$\\ 
  & $\boldsymbol{\alpha}$ & $\boldsymbol{99.46\%}$ & $\boldsymbol{90.54\%}$ \\
 & $\alpha_{+}$ & $99.57\%$ & $86.58\%$  \\
 \hline \vspace{-0.3cm}
 \multirow{3}{*}{SALI}  &  &  & \\
  & $\alpha_{-}$ & $99.05\%$ & $91.61\%$ \\ 
  & $\boldsymbol{\alpha}$ & $\boldsymbol{99.32\%}$ & $\boldsymbol{90.71\%}$ \\
 & $\alpha_{+}$ & $99.49\%$ & $86.8 \%$  \\
 \hline  \vspace{-0.3cm}
 \multirow{3}{*}{FLI}  &  &  &     \\
  & $\alpha_{-}$ & $97.95\%$ & $91.86\%$ \\ 
  & $\boldsymbol{\alpha}$ & $\boldsymbol{98.28\%}$ & $\boldsymbol{91.43\%}$ \\
 & $\alpha_{+}$ & $98.54\%$ & $87.73\%$ \\
 \hline\hline

 \multicolumn{4}{c}{$k=3.0$ }\\

 \hline \hline \vspace{-0.3cm}
 \multirow{3}{*}{$\overline{Y}$}  &  &  &      \\
  & $\alpha_{-}$ & $99.98\%$ & $99.73\%$ \\ 
  & $\boldsymbol{\alpha}$ & $\boldsymbol{99.98\%}$ & $\boldsymbol{99.73\%}$ \\
 & $\alpha_{+}$ & $99.99\%$ & $99.72\%$ \\
 \hline \vspace{-0.3cm}
 \multirow{3}{*}{SALI}  &  &  & \\
  & $\alpha_{-}$ & $99.98\%$ & $99.73$\\ 
  & $\boldsymbol{\alpha}$ & $\boldsymbol{99.98\%}$ & $\boldsymbol{99.73\%}$ \\
 & $\alpha_{+}$ & $99.98\%$ & $99.72\%$  \\
 \hline  \vspace{-0.3cm}
 \multirow{3}{*}{FLI}  &  &  &     \\
  & $\alpha_{-}$ & $99.97\%$ & $99.73\%$\\ 
  & $\boldsymbol{\alpha}$ & $\boldsymbol{99.97\%}$ & $\boldsymbol{99.73\%}$ \\
 & $\alpha_{+}$ & $99.98\%$ &  $99.72\%$ \\
 \hline

\end{tabular}

\end{tabular}
\caption{Percentages of agreement between the labels produced by $\dldda$ (left) and $\dldfd$ (right) and the ones produced by mean MEGNO, SALI, and FLI, for the values of the thresholds depicted in Fig.\ \ref{fig:DeltaLDThresholdsPicture}. See text for more details.} 
\label{table:DeltaLDThresholdsTable}
\end{table}

%\clearpage 

\bibliographystyle{apalike} 
\bibliography{biblio}

\end{document}